\documentclass[final,3p,times,twocolumn]{elsarticle}
 \biboptions{comma,sort&compress}
\usepackage{here}
 \usepackage{graphicx}
  \usepackage{epsfig}

\def\nin{\noindent}
\def\beq{\begin{equation}}
\def\eeq{\end{equation}}
\def\bea{\begin{eqnarray}}
\def\eea{\end{eqnarray}}

\journal{Nuc. Phys. (Proc. Suppl.)}

\begin{document}

\begin{frontmatter}



\title{Precise determination of the
 $f_0(500)$ and $f_0(980)$ parameters 
  in dispersive analysis of the $\pi\pi$ data}

 \author[label1]{Robert Kami\'nski, \corref{cor1}}
  \address[label1]{Institute of Nuclear Physics PAN, Krak\'ow}
\cortext[cor1]{Speaker}
\ead{robert.kaminski@ifj.edu.pl}

 \author[label2]{R. Garcia-Martin}
 \author[label2]{J. R. Pelaez}
 \author[label2]{J. Ruiz de Elvira}
 \address[label2]{Dept. F\'{\i}sica Te\'orica II. Universidad Complutense. 28040 Madrid, SPAIN}


\begin{abstract}
\noindent

We review the use of new and precise dispersive equations,
which also implement crossing symmetry, in order to shed further light on the long-standing puzzle in the parameters of the $f_0(500)$, as well as the $f_0(980)$.
This puzzle is finally being settled thanks to several
analyses carried out during the last years \cite{PDG2012}. 
In this talk we show how our very recent dispersive data analysis allowed
for a precise and model independent determination of the amplitudes
for the $S,P,D$ and $F$ waves \cite{GKPY2011A,GKPY2011B,GKPYDF}. 
In particular, we show how the analytic continuation of once subtracted dispersion relations for the $S0$ wave to the complex energy plane leads to very precise results for the $f_0(500)$ pole: $\sqrt{s}_{pole} = 457^{+14}_{-13} - i 279^{+11}_{-7}$ MeV and for the $f_0(980)$ pole: $\sqrt{s}_{pole} = 996\pm 7  - i 25^{+10}_{-6}$~MeV.
We also comment on how these results have been
already used for other practical applications,
including a refit of a previous model to the
$\pi\pi$ $S$-wave amplitudes below 1000 MeV, which improves its consistency
with the poles found with the dispersive approach.

\end{abstract}





\end{frontmatter}


\section{Introduction}
\nin
We have recently constructed a set of $\pi\pi$ scattering $S, P, D$ and $F$ partial waves 
 \cite{GKPY2011A,GKPY2011B,GKPYDF} as simple and ready to use fits to data covering
the energy range from the threshold to about 1400 MeV 
and including very recent $K_{l4}$ experimental results \cite{Batley:2010zza}. The relevant observation is that
stringent theoretical constraints from Roy-type dispersion relations, forward dispersion relations (FDR) and sum rules for threshold parameters \cite{GKPY2011A} have been added to the initial unconstrained fit to the data.
In particular, we have imposed that the $\pi\pi$ input
 $S, P, D$ and $F$ partial waves should agree within uncertainties with the output of the dispersion relations, while still reproducing the data. Hence, despite their simplicity, these constrained fits are very consistent, reliable and precise.

This consistency of all input partial waves expressed with dispersion relations ensures that
the  analytic continuation to the complex plane, by means of the very same dispersive integrals, provides reliable and precise information, which is particularly interesting for the $S$-wave and the $f_0(500)$ and $f_0(980)$ resonances.
The high precision attained just from a data analysis has been achieved due to the implementation of very demanding once subtracted dispersion relations respecting crossing symmetry (so called GKPY equations \cite{GKPY2011A}). We briefly report here on the amplitudes and the results of their extrapolation to the complex plane.

\section{Method and results}
\nin

The GKPY dispersion relations, are Roy-type equations \cite{Roy:1971tc}
that can be recast as relations between the real and imaginary parts of the output and input partial waves $t_{\ell}^{I}(s=m_{\pi\pi}^2)$, namely: 
\begin{eqnarray}
\hspace{-.7cm}{\mbox{Re } t_{\ell}^{I{(OUT)}}(s)} &\hspace{-.2cm} = \hspace{-.2cm}& \sum\limits_{I'=0 }^2C_{st}^{II'}{a_0^{I'}} \\
\hspace{-.7cm}&\hspace{-.2cm} + \hspace{-.2cm}&  
\displaystyle \sum\limits_{I'=0}^2
        \displaystyle \sum\limits_{\ell'=0}^4
     \hspace{0.15cm}-\hspace{-0.37cm}
        \displaystyle \int \limits_{4m_{\pi}^2}^{\infty}\ \hspace{-.1cm}ds'
     K_{\ell \ell^\prime}^{I I^\prime}(s,s')\, { \mbox{Im }t_{\ell'}^{I^{\prime{(IN)}}}(s')}
\end{eqnarray}     
where the first terms on the right are called subtraction terms, which are combinations of the   $S$ wave scattering lengths ${a_0^{I'}}$. The $K_{\ell \ell^\prime}^{I I^\prime}(s,s')$ are kernels derived  by imposing crossing symmetry conditions .

Our fits are obtained by minimizing the difference $\mid{\mbox{Re } t_{\ell}^{I{(OUT)}}(s)} - {\mbox{Re } t_{\ell}^{I{(IN)}}(s)}\mid$ which can be treated as measurement of 
fulfilment of crossing symmetry by analyzed amplitudes.
The smaller this difference the better analyticity and crossing symmetry  are satisfied.

Actually, a consistency check of the data fits with all theoretical constraints has been done by minimization of the function 
$\chi_{tot}^2 = \chi^2_{data} + \bar d^2_{Roy} + \bar d^2_{GKPY} + \bar d^2_{FDR} + \bar d^2_{SR}$
where $\bar d_i^2 = \frac{1}{np} \sum\limits_j^{np} \left(\frac{\Delta_i(s_j)}{\delta\Delta_i(s_j)}\right)^2$ are averaged distances between input and output real parts of amplitudes. 
The  OUT-IN difference for a given dispersion relation at an energy $\sqrt{s_i}$ is denoted by 
$\Delta_i(s_j)$ and its uncertainty by $\delta\Delta_i(s_j)$.
The final values of these averaged differences for the Roy and GKPY equations and FDR were:
$\bar d^2_{Roy} = 0.14$, 
$\bar d^2_{GKPY} = 0.32$ and
$\bar d^2_{FDR} = 0.4$.
This is a remarkable fulfillment of analytic and crossing constraints.
Note that GKPY equations are still very well satisfied even though their
$\delta\Delta_i(s_j)$ are much smaller than for Roy equations above 400 MeV, given the same input.
Let us remark that when we fit only to experimental data without imposing the dispersion relations within uncertainties, these numbers are 0.87, 1.9 and 1.98, respectively.

\begin{figure}
  \begin{center}
  \includegraphics[width=1\columnwidth,height=18cm]{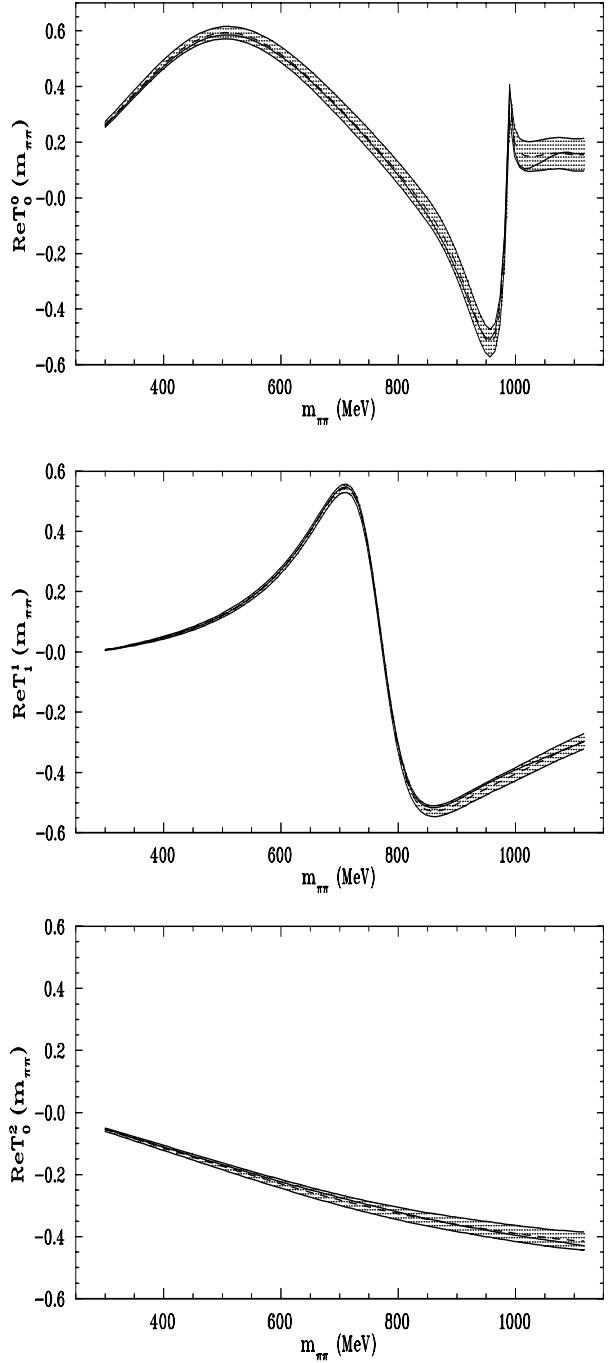}
\end{center}
\caption{Real parts of the input (dashed lines) and output (solid lines) amplitudes for the GKPY equations together with the uncertainties in the difference between both of them (gray bands). From top to bottom: for the $S0$ wave, $P$ wave and $S2$ wave.}
\label{fig:DT}       
\end{figure}


The curves on Fig. \ref{fig:DT} compare the input and output for the GKPY equations. One can see that the $\Delta_i(s_j)$ differences are everywhere smaller than $\delta\Delta_i(s_j)$.
For our purposes, the most relevant feature of GKPY equations is the 
very slowly increasing uncertainties, due to the fact that, contrary to Roy equation case, the GKPY subtraction terms are constant and the uncertainties they produce do not grow with increasing energy (see discussion in \cite{GKPY2011A}).
Note that, of the infinite tower of Roy and GKPY equations, our analysis only uses up
to the $P$ wave equations.
The $D$ and $F$ waves are therefore only constrained by their role as input in the $S,P$ 
equations, but not directly in dispersion relations for the $D$ or $F$  waves. A later study \cite{GKPYDF} of $D$ and $F$ GKPY equations
also indicates a fairly good consistency.

\begin{figure}
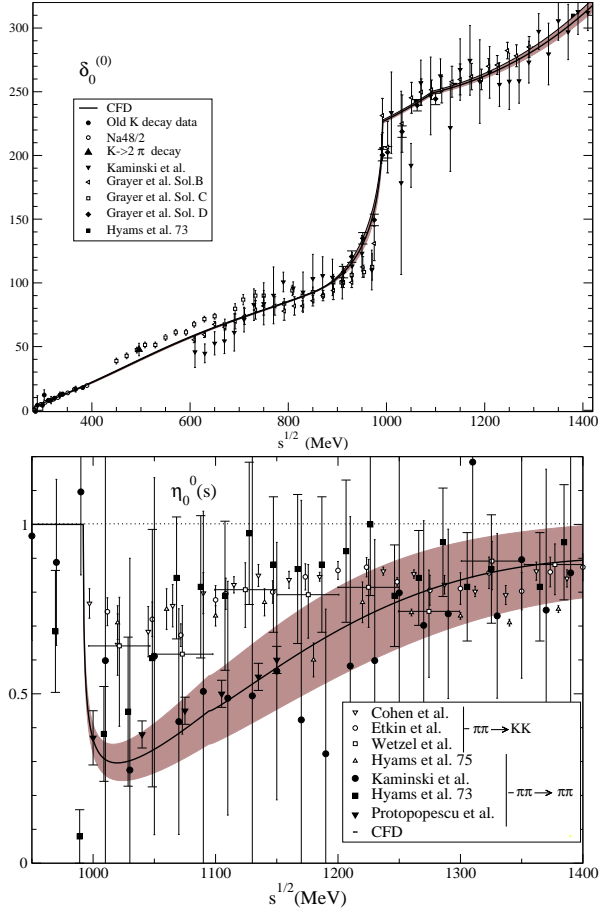

\resizebox{1.\columnwidth}{!}{
\includegraphics{CFD-S0wave.eps}}

\resizebox{1.0\columnwidth}{!}{
\includegraphics{CFD-S0inel.eps}}

\caption{Phase shifts and inelasticities together with their uncertainties (gray bands) for the $S0$ wave calculated in the fit to the data (for details see \cite{GKPY2011A}) constrained with dispersion relations.}
\label{fig:phaseeta}       
\end{figure}

On Fig. \ref{fig:phaseeta} the phase shifts and inelasticities for the $S0$ wave amplitude are presented.
They represent results of our fit to experimental 
data constrained to dispersion relations, namely, the
so called Constrained Fit To Data (CFD). 
Two kinks on the phase shifts around 992 MeV and 1100 MeV are produced by the opening of two new thresholds: $K\bar K$ and $\eta\eta$.
On the figure for the inelasticity one can easily see that our fit strongly prefers a solution
with a "dip" around $1000-1100$~MeV.
Such behavior of inelasticities is, in general, more compatible with the $\pi\pi \to \pi\pi$ 
data sets than with the $\pi\pi\rightarrow \bar KK$ ones.

Analytic continuation to the energy complex plane 
of the output amplitudes from Roy and GKPY equations 
yields poles related to the $f_0(500)$, $f_0(980)$ and $\rho(770)$ resonances and allows for a 
calculation of their couplings to the $\pi\pi$ channel. 
The parameters of these poles and couplings are presented in Table \ref{tab:polesCFD}. 
The couplings, when obtained from the residues of the poles, are normalized as: 
\begin{equation}
g^2=-16\pi \!\!\lim_{s\rightarrow s_{pole}}\!\!(s-s_{pole})\,t_{\ell}(s)\,(2\ell+1)/(2p)^{2\ell} \nonumber 
\end{equation}
where $p^2=s/4-m_\pi^2$.
One can see that the central values of all
parameters and couplings either from Roy or GKPY equations differ
only slightly and always lie within the estimated uncertainties.
This confirms the compatibility of the twice and once subtracted dispersion relations used in our fits. 
Let us emphasize that
the uncertainties are typically smaller for 
GKPY results than for those coming from Roy equations.

\begin{table}
\begin{center} 
\begin{tabular}{|c|c|c|}
\hline
&$\sqrt{s_{\rm{pole}}}$ (MeV)&$\vert g \vert$\\\hline
$f_0(500)^{\rm GKPY}$ & {$(457^{+14}_{-13})-i(279^{+11}_{-7})$}& {$3.59^{+0.11}_{-0.13}$} GeV\\
$f_0(500)^{\rm Roy}$ & $(445\pm25)-i(278^{+22}_{-18})$&$3.4\pm0.5$ GeV\\
\hline
$f_0(980)^{\rm GKPY}$ & {$(996\pm7)-i(25^{+10}_{-6})$}&{$2.3\pm0.2$} GeV\\
$f_0(980)^{\rm Roy}$ & $(1003^{+5}_{-27})-i(21^{+10}_{-8})$ &$2.5^{+0.2}_{-0.6}$ GeV\\
\hline
$\rho(770)^{\rm GKPY}$ & {$(763.7^{+1.7}_{-1.5})-i(73.2^{+1.0}_{-1.1})$}&{$6.01^{+0.04}_{-0.07}$}\\
$\rho(770)^{\rm Roy}$ & $(761^{+4}_{-3})-i(71.7^{+1.9}_{-2.3})$& $5.95^{+0.12}_{-0.08}$\\
\hline
\end{tabular}
\end{center}
\caption{Poles and residues from Roy and GKPY equations.}
\label{tab:polesCFD}
\end{table}

\begin{figure}
\resizebox{1.0\columnwidth}{!}{
\rotatebox{90}{
\includegraphics{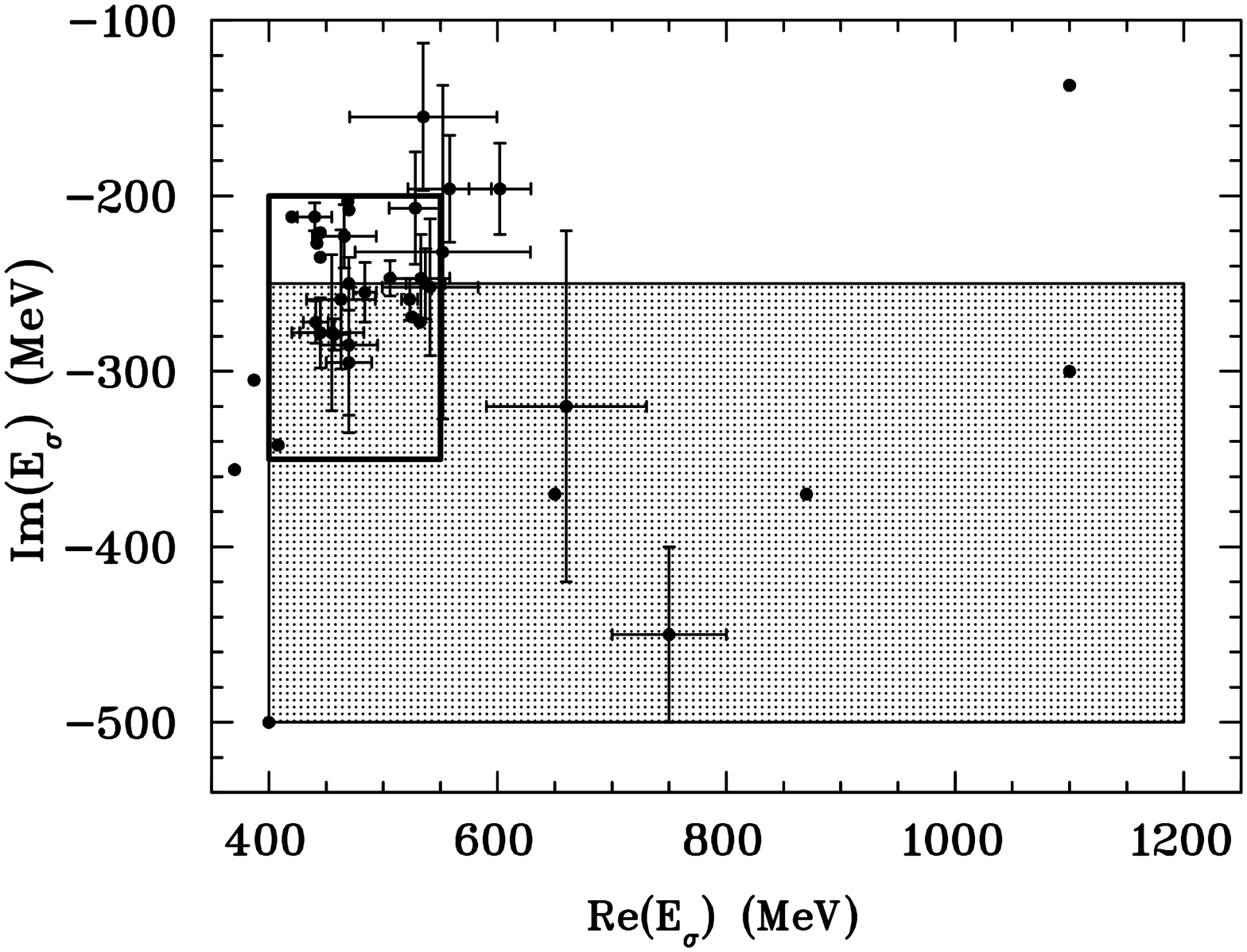}}}
\resizebox{1.0\columnwidth}{!}{
\rotatebox{90}{
\includegraphics{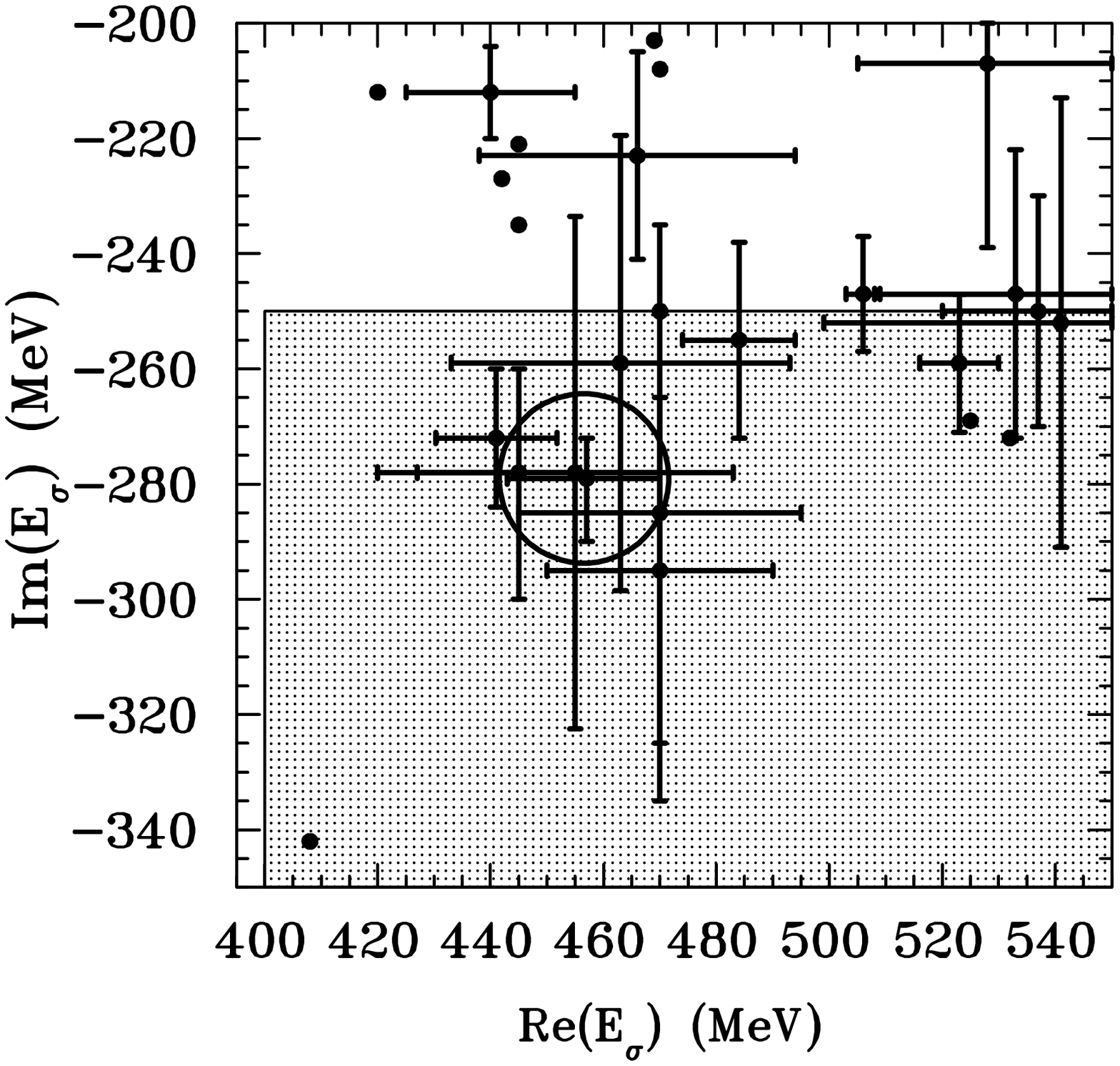}}}
\caption{Positions of the poles related with $f_0(500)$ (or $\sigma$) cited in PDG'2010 
\cite{PDG2010} - black dots. 
Gray bands represent uncertainties  in the  mass and minus half of the width of the $f_0(500)$ in PDG'2010. 
The rectangle on the upper panel indicates the magnified area shown on the lower panel and corresponds to the 
errors of mass and minus half of the width of the $f_0(500)$ in PDG'2012 \cite{PDG2012}.
The pole calculated in this work lies well inside this improve uncertainty region. Here $E_{\sigma} = \sqrt{s_{\sigma}}$}
\label{fig:sigma}       
\end{figure}

On Fig. \ref{fig:sigma} we show the position
of the pole found in this analysis and of different
``$f_0(600)$'' poles taken from the compilation made in 2010 by the Particle Data Group (PDG) \cite{PDG2010}. However, shortly after the presentation of this talk, the uncertainties, estimated by the PDG in its latest 2012 edition, have shrunk dramatically, in part due to their considering the result of this work.
For comparison, 
 we show as a rectangle this new estimation for the mass and width
in the 2012 PDG\cite{PDG2012}, where even the name of the particle has been changed to $f_0(500)$, which is the one we have used in this text. Moreover, as already pointed out in \cite{GKPY2011A,GKPY2011B},
our results agree very well 
with those obtained in \cite{CCL2006}, where Roy equations were used 
together with Chiral Perturbation Theory (ChPT) predictions to obtain a precise determination of the $f_0(500)$ pole. Our results do not use 
ChPT and the nice agreement with \cite{CCL2006} is therefore a good consistency check for ChPT too.

One can easily notice the significant improvement
in the estimated uncertainties:
The mass has changed from M = $400-1200$ MeV to $400-550$ MeV and the width  from 
$\Gamma = 2\times (250-500)$ MeV to $2\times (200-350)$ MeV.
The result we have reviewed here has also influenced the estimated
values in the PDG 2012 for the next light scalar-isoscalar 
state, i.e.: the $f_0(980)$, which are: $M_{f_0(980)} = 990 \pm 20$ MeV and
$\Gamma_{f_0(980)} = 40-100$ MeV.
The position of the pole found in this analysis is 
$M_{f_0(980)} = 996 \pm 7$ MeV and
$\Gamma_{f_0(980)} = 50^{+20}_{-12}$ MeV.

\section{Application}
\nin

As we have commented above, the most remarkable features 
of the parameterizations we have obtained is their precision and consistency, but we would also like to remark
that they are rather simple and ready to use. Actually, they have already been used by several groups
as input in several  precision analysis with pions in the final state (see \cite{Schneider:2012ez} and earlier references therein).

Another interesting example of practical application 
of the method described above has been presented by one of us
and other collaborators in the International IUPAP Conference on Few-Body Problems in Physics \cite{FB2012}.
The authors modified their older model \cite{Surovtsev}, constructed by fits only to experimental data in the range from the $\pi\pi$ threshold to about 1800 MeV, by adding the GKPY equations. 
The amplitude below about 1000 MeV has changed dramatically. 
For example the older sigma pole at $616.5 - i 554.0$ MeV moved to $473.7 - i 297.8$ MeV i.e. to the position located less than one standard deviation from that presented here.
The total $\chi2$ was changed from almost 2.4 to 1.3.
In addition, the threshold parameters thus obtained are
consistent with \cite{GKPY2011A}.

\section{Conclusions}
\nin
The method presented here of combining the analysis of experimental data with theoretical constraints in the form of dispersion relations 
is model independent, very efficient, precise and yields parameterizations which are very easy to use later on.
In addition, it provides a reliable extrapolation to the complex plane, which has allowed us to obtained a precise determination of the
controversial poles that appear at low energies in the scalar-isoscalar wave.
Actually, it seems that the relatively recent and
precise dispersive determination of the 
parameters of the two lightest scalar-isoscalar 
mesons $f_0(500)$ and $f_0(980)$s, 
and in particular the one we are reviewing here, 
have finally lead to a significant reevaluation of the estimates
in the new edition of the particle data tables 
\cite{PDG2012}.
It should be noted that even 
the name of the$f_0(500)$ has changed (previously it was $f_0(600)$). 

We have also shown examples of how the 
use of GKPY equations can dramatically modify
existing models when analyzing
$\pi\pi$ amplitudes and improve their agreement 
with crossing symmetry.

We hope that our method and our amplitudes can 
be of help for further studies in Hadronic Physics by improving
their treatment of $\pi\pi$ interactions. 

\section*{Acknowledgements}
\nin
This work has been partly supported by the Polish Ministry of Science and
Higher Education (grant No N N202 101 368) as well as by DGICYT contracts FPA2011-27853-C02-02, 
FPA2010-17806, and the EU Integrated
Infrastructure Initiative Hadron Physics Project under contract
RII3-CT-2004-506078.










\end{document}